\documentclass[a4paper,fleqn]{cas-dc}

\usepackage[numbers]{natbib}
\usepackage{color,soul}     %for highlight
\usepackage[version=4]{mhchem}

\def\tsc#1{\csdef{#1}{\textsc{\lowercase{#1}}\xspace}}
\tsc{WGM}
\tsc{QE}
\newcommand{\eg}{e.\,g.}

\newcommand{\rcm}{\ensuremath{\text{cm}^{-1}}}
\newcommand{\cm}{\ensuremath{\text{cm}^{-1}}}

\newcommand{\kJmol}{\ensuremath{\text{kJ}\,\text{mol}^{-1}}}
\newcommand{\massU}{\ensuremath{m/z}}

\newcommand{\kmmol}{\ensuremath{\text{km} \cdot \text{mol}^{-1}}}

\newcommand{\HCNHp}{\ce{HCNH+}}

\begin{document}
\let\WriteBookmarks\relax
\def\floatpagepagefraction{1}
\def\textpagefraction{.001}

\shorttitle{}    

\shortauthors{}  

\title [mode = title]{Rovibrational Overtone and Combination Bands of the \HCNHp\ Ion}

\author[1]{Miroslava Kassayov{\' a}}%[<options>]

\affiliation[1]{organization={Department of Surface and Plasma Science, Faculty of Mathematics and Physics, Charles University},
            addressline={V Hole{\v s}ovi{\v c}k{\' a}ch 2}, 
            city={Prague},
            postcode={18000},
            country={Czech republic}}

\author[2]{Miguel Jiménez-Redondo}%[]

\affiliation[2]{organization={Max Planck Institute for Extraterrestrial Physics},
            addressline={Gießenbachstraße 1}, 
            city={Garching},
            postcode={85748}, 
            country={Germany}}

\author[3,4]{J{\' a}nos Sarka}%[]

\affiliation[3]{organization={I. Physikalisches Institut, Universität zu Köln},
            addressline={Zülpicher Str. 77}, 
            city={Cologne},
%          citysep={}, % Uncomment if no comma needed between city and postcode
            postcode={50937}, 
            country={Germany}}
\affiliation[4]{organization={Institute of Chemistry, Eötvös Loránd University},
            addressline={Pázmány Péter sétány 1/A.}, 
            city={Budapest},
            postcode={1117}, 
            country={Hungary}}

\author[1]{Petr Dohnal}%[]

\author[1]{Juraj Glos{\' i}k}%[]

\author[2]{Paola Caselli}%[]

\author[2]{Pavol Jusko}%[]
\cormark[1]
%% Footnote of the second author

% Email id of the second author
\ead{pjusko@mpe.mpg.de}

\cortext[1]{Corresponding author}

\begin{abstract}
Spectra of vibrational overtone and combination bands from vibrational ground state of \HCNHp\ 
were measured using an action spectroscopy technique with active background suppression 
in a cryogenic 22 pole radio frequency ion trap apparatus. 
Spectroscopic constants for the upper vibrational levels  of the transitions were determined with 
vibrational band origins being 6846.77981(90) \rcm\ ($2\nu_1$, NH stretch), 
6640.47624(43) \rcm\ ($\nu_1+\nu_2$), 6282.03578(63) \rcm\ ($2\nu_2$, CH stretch), and 
6588.4894(20) \rcm\ ($\nu_2+\nu_3+2\nu_5^0$). 
State of the art \emph{ab initio} VCI calculations
up to $10^4\;\rcm$ complement the experimental data.   
\end{abstract}

% Keywords

\maketitle

\section{Introduction}\label{}

The gas-phase ion-molecule chemistry of both diffuse and dense interstellar clouds has been extensively investigated over 
the years, contributing to our understanding of the interstellar medium (ISM) and its complex chemical processes \cite{Millar2015}. 
Among the key molecular ions studied is \HCNHp, a linear, closed-shell species that plays a significant role in  
interstellar chemistry \cite{Herbst1973,Quenard2017,Fontani2021,Dohnal2023}. \HCNHp\ is considered to be the primary 
precursor for the formation of HCN and HNC \cite{Semaniak2001,Quenard2017,Fontani2021}, both of which are crucial 
molecules in the ISM \cite{Loison2014}.
In particular, \ce{HCN} is considered an essential precursor of prebiotic molecules, 
including sugars, nucleotides, amino acids, and fatty acids (\eg \cite{Todd2020,Pearce2022}). 
The HCN-to-HNC intensity ratio also provides information on the kinetic temperature of interstellar 
molecular clouds \cite{Hacar2020}.

The presence of \HCNHp\ has been confirmed in various interstellar regions, including the dense and
cold dark cloud TMC-1 in the Taurus Molecular Cloud complex \cite{Ziurys1992}, the pre-stellar core L1544 \cite{Quenard2017}, 
high-mass star forming regions \cite{Fontani2021} and
the molecular cloud Sgr B2, where its rotational transitions have been 
originally detected \cite{Ziurys1986}.

Due to its astrophysical relevance, \HCNHp\ has been the subject of numerous experimental and theoretical studies. 
In the laboratory, this ion has been characterized through rotationally resolved infrared spectroscopy and further 
studied across a wide spectral range, from microwave \cite{Araki1998} to sub-millimeter wavelengths \cite{Bogey1985,Amano2006,Silva2024}. 
Fundamental vibrational bands of \HCNHp\ have been previously observed: $\nu_1$ (NH stretch) \cite{Altman1984}, 
$\nu_2$ (CH stretch) \cite{Altman1984b}, $\nu_3$ (CN stretch) \cite{Kajita1988,Liu1988}, 
$\nu_4$ (HCN bend) \cite{Tanaka1986} and $\nu_5$ (HNC bend) \cite{Ho1987}. 
Additionally, the $\nu_1+\nu_4 \leftarrow \nu_4$ and $\nu_1+\nu_5 \leftarrow \nu_5$ hot bands 
were also explored \cite{Amano1986}. 
Alongside the experiments, several theoretical studies 
\cite{Botschwina1986, Botschwina1994,Peterson1995,Brites2012,Cotton2013} 
were focused on the determination of the structure and spectroscopic properties of \HCNHp\ ions. 

Experimentally determined frequencies of transitions involving excitation of multiple vibrational quanta 
reveal information about the molecule's potential energy surface (PES) and as such are very useful to benchmark \emph{ab initio} calculations.
To our best knowledge, the overtone transitions of \HCNHp\ ions have not yet been experimentally studied. 
This paper reports the observation of the rotationally resolved $2\nu_1$ and $2\nu_2$ overtone and $\nu_1+\nu_2$ and $\nu_2+\nu_3+2\nu_5^0$ 
combination bands of the \HCNHp\ ion.

\section{Material and Methods}\label{}

\subsection{Experimental}

The experimental setup, Cold CAS Ion Trap -- a cryogenic 22 pole radio frequency (rf) trap, has been 
extensively described previously \cite{Jusko2023} 
and only a short description will follow. The ion of interest, \HCNHp, was produced in an ion storage source using electron 
bombardment in a mixture of \ce{HCN}, He, and \ce{H2}. The ion beam exiting the source is mass selected using a quadrupole mass filter and injected 
into a 22 pole ion trap mounted on top of a closed cycle helium cryostat. Variable temperature is achieved using a resistive heater.
A short and intense helium pulse is used to trap and cool the injected ions close to the temperature of the trap walls. 
After a storage time of several seconds (typically $2.5\;\text{s}$), 
the trap is emptied towards a product quadrupole mass filter and the ions are counted in a Daly type detector.
Laser light was delivered by an Agilent 8164B option 200 light system 
($1440-1640\;\text{nm}$, line width $<100\;\text{kHz}$, and, 
power $1-8\;\text{mW}$ (wavelength dependent)). 
The wavelength was measured by an EXFO WA 1650 wavemeter 
with an absolute accuracy better than 0.3 ppm at 1500 nm.          
The scanning has only been performed in the vicinity of selected transitions and the scanning window
got progressively narrower as the fit precision increased with the number of detected lines for each band.
Two kodial glass windows aligned with the trap axis enabled us to use a triple pass 
laser beam path configuration \cite{Jimenez2024}. 

In the present experiments, a variation of a laser induced reaction (LIR) technique \cite{Schlemmer1999} (an action spectroscopy scheme),
was applied in order to  obtain overtone spectra of \HCNHp\ ions. 
The LIR technique takes advantage of a different reactivity of the ground and excited 
states of the ion of interest. For this purpose, reactant gas \ce{C2H4}, ethylene, has been continuously leaked into the trap
(effective number density in the trap in the $10^{11}\,\text{cm}^{-3}$ range) at $T=125\;\text{K}$ to avoid excessive freezing 
and the following reactions took place:
\begin{align}
\label{e_HCNHpC2H4}
\ce{HCNH+} + \ce{C2H4} &\to \ce{C2H5+} + \ce{HCN}, && \Delta H=32.4~\kJmol \\   
\label{e_C2H5pC2H4}
\ce{C2H5+} + \ce{C2H4} &\to \ce{C3H5+} + \ce{CH4}, && \Delta H < 0         \\
\label{e_C3H5pC2H4}
\ce{C3H5+} + \ce{C2H4} &\to \ce{C5H7+} + \ce{H2},  && \Delta H < 0
\end{align}
where the endothermicity of reaction (\ref{e_HCNHpC2H4}) was calculated from the experimental proton affinities taken 
from the NIST database \cite{Lindstrom2023}. 
Reaction (\ref{e_HCNHpC2H4}) is the only endothermic reaction in this system and the exothermic reactions (\ref{e_C2H5pC2H4}--\ref{e_C3H5pC2H4})
proceed with close to Langevin collisional reaction rate, as has been tested in a preparatory experiment, where \ce{C2H5+}
has been injected into the trap directly.
The final product in this reaction scheme, \ce{C5H7+}, reacts only very slowly with \ce{C2H4}, and is thus monitored as
a proxy for the \HCNHp\ photon excitation. We will further refer to this process as a ``multi-step LIR'' scheme. 
Moreover, as the $m/z$ of the ion of interest \HCNHp\ ($28\;\massU$) is more than two times lower than that of the monitored ion 
\ce{C5H7+} ($67\;\massU$), we also applied the zero background ``kick-out'' technique \cite{Jimenez2024}, where the higher \massU\ product 
ions contributing to the background are 
removed from the trap by lowering the rf amplitude for few hundred milliseconds, 
shortly before the irradiation of \HCNHp\ takes place.  
This ensures that all the \ce{C5H7+} product ions are formed throughout the irradiation time from the 
LIR scheme (\ref{e_HCNHpC2H4})--(\ref{e_C3H5pC2H4}).

\subsection{Computational}\label{Computational}
The anharmonic vibrational frequencies computed in this study
were determined using the vibrational configuration interaction theory
(VCI) \cite{21MaRa,21MaPeRa,24ScRa}
implemented in the MOLPRO package of ab initio programs \cite{molpro}.
First, the equilibrium geometry of \HCNHp\ has been determined
using explicitly correlated coupled-cluster theory
including single and double excitations and
a perturbative treatment of the triple excitations \cite{07AsKnWe}
with an augmented correlation-consistent basis set \cite{08PeAdWe}
with the core electrons being also correlated \cite{10HiMaPe},
CCSD(T)-F12b/cc-pCVQZ-F12.
Second, a harmonic frequency calculation was carried out
in order to determine displacement vectors,
which were used to span the multidimensional PES
with the equilibrium geometry serving as the reference point for the expansion.

Next, an $n$-mode expansion \cite{08BoCaMe}
being truncated after the 4-mode coupling terms was 
utilized to generate the PES.
A multilevel scheme was used \cite{24ScRa},
where the first and second order terms
were computed at the level indicated above,
while a smaller basis set, cc-pCVTZ-F12,
was used for the 3rd and 4th order terms.
The PES was then transformed from a grid representation
to an analytical representation \cite{16ZiRa}.
Using the $n$-mode polynomial PES, one-mode wavefunctions (modals)
were determined first by vibrational self-consistent field (VSCF) \cite{19MeHaKaRa}.
These modals served as basis functions for the following VCI calculations.
Vibrational states were computed up to 4 quanta excitations below 10000 \rcm.
The VCI configuration space converged with
the maximum number of simultaneously excited modes of seven,
the maximal excitation level per mode of twelve, and 
the maximal sum of quantum numbers of 15.
The VCI convergence errors were below 0.1 \cm,
while the final size of the VCI matrix was 170,488.

\section{Results and Discussion}\label{}

\begin{figure}%[]
  \centering
    \includegraphics{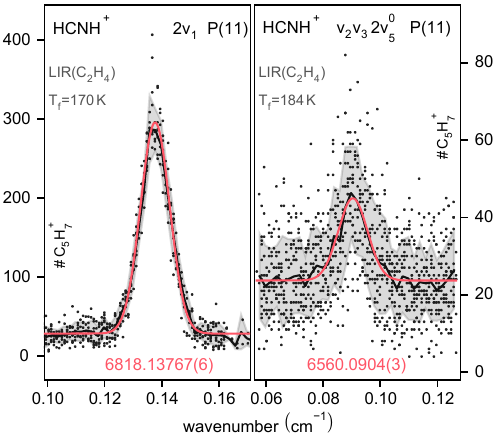}
    \caption{Measured absorption line profiles for P(11) transitions belonging to the $2\nu_1$ 
    (left panel) and $\nu_2+\nu_3+2\nu_5^0$ (right panel) vibrational bands of \HCNHp\ ions.
    The abscissa is offset by 6818$\;\rcm$ (left panel) and 6560$\;\rcm$ (right panel).
    The temperatures $T_f$ listed in the figure were obtained from the Doppler broadening of the absorption lines. 
    Nominal trap temperature was 128~K.    
    }\label{fig1}
\end{figure}

Two selected absorption lines profiles of the \HCNHp\ ion acquired using the multi-step LIR scheme are shown in Figure~\ref{fig1}. 
The frequency dependent numbers of trapped \ce{C5H7+} ions, mimicking the absorption in the \HCNHp\ ion, were fitted by a Doppler 
profile. The resulting center transition wavenumbers are summarised in Table~\ref{t_lines}. 
To describe the upper and lower energy levels of \HCNHp\, we used a standard Hamiltonian for a linear molecule \cite{Jimenez2024}:
\begin{equation}
\label{Hamiltonian}
H = T_\nu+B_\nu J (J+1)-D_\nu [J(J+1)]^2+H_\nu [J(J+1)]^3,
\end{equation}
where $\nu$ represents vibrational quantum numbers of a given state, $T_\nu$, $B_\nu$, $D_\nu$ and $H_\nu$ are 
spectroscopic constants and $J$ is the rotational quantum number. 
As all the probed vibrational bands originated in the vibrational ground state, $T_\nu$ corresponds to the band origin. 
The measured transition wavenumbers were fitted using Hamiltonian (\ref{Hamiltonian}) while the spectroscopic 
constants for the lower state were fixed at the values recently reported by \citet{Silva2024}. 
The obtained spectroscopic constants for the measured overtone and combination bands of \HCNHp\ are 
listed in Table \ref{t_constants} together with the present theoretical predictions for the band 
origins ($T_\mathrm{calc}$) and infrared intensities ($I_\mathrm{calc}$). 
For the three strongest bands, $T_\mathrm{calc}$ is within $7\;\rcm$ of the experimentally determined values.   

\begin{table*}[width=.95\textwidth,cols=5,pos=h]
    \caption[]{Spectroscopic constants of the upper levels of the vibrational bands of \HCNHp\ probed in the present study. 
    $T_\mathrm{calc}$ denotes the predicted vibrational band origins and $I_\mathrm{calc}$ the calculated infrared intensities.}\label{t_constants}
    \begin{tabular*}{\tblwidth}{@{} CCCLCCC@{} }
        \toprule
        Band & $I_\mathrm{calc}$ (\kmmol)   & ~$T_\mathrm{calc}$ (\rcm)     &  ~~~$T_\nu$ (\rcm)~~~  & $B_\nu$ (\rcm)              
            &  $D_\nu$  ($10^{-6}$ \rcm)  & $H_\nu$  ($10^{-9}$ \rcm)       \\
        \midrule
        $2\nu_1$               & 1.92  & 6851.15  &  6846.77981(90)    & 1.222953(37)  & 3.34(42)   & 2.9(1.3)          \\
        $\nu_1+\nu_2$          & 1.98  & 6646.76  &  6640.47624(43)    & 1.221567(13)  & 1.698(72)  &          \\
        $2\nu_2$               & 0.44  & 6284.64  &  6282.03578(63)    & 1.220617(34)  & 1.08(39)   &         \\
        $\nu_2+\nu_3+2\nu_5^0$ & 0.06  & 6608.61  &  6588.4894(20)     & 1.225412(62)  & 5.21(43)   &          \\
        \noalign{\smallskip}
        \bottomrule
    \end{tabular*}
    \vskip 0.2em
    {\footnotesize{\emph{Note:} 
    Numbers in parentheses denote the statistical error of the fitted parameters in the units of the last quoted digit.
    $T_\mathrm{calc}$ and $I_\mathrm{calc}$ were computed with a hybrid VCI calculation at the
    AE-CCSD(T)/cc-pCVQZ-F12//AE-CCSD(T)/cc-pCVTZ-F12 level of theory. For details see section \ref{Computational}.  
    }}
\end{table*}

A weak band was observed close to the $\nu_1+\nu_2$ combination band.
Based on the calculated band origins and vibrational transition moments, 
as described in section \ref{Computational}, we identified this band as $\nu_2+\nu_3+2\nu_5^0$. 
Note that the value of the calculated band origin for this band is 20 \rcm\ higher than the experimental one.
However, there are no other states within ~$\pm\,150\;\cm$ apart from the 
transition to $\nu_2+\nu_3+2\nu_5^2$ at $6621.21\;\cm$, which is not allowed due to the selection 
rules ($\Delta l=0,\pm 1$ for a linear molecule).
It is very challenging to infer absolute transition intensities from action spectroscopy 
spectra \cite{Jimenez2024}, and even relative values have to be taken with caution as 
the overlap between the ion cloud in the trap and the laser beam can change during experiments, 
especially with changing number of trapped ions
(for illustration see the full line list and scattered Boltzmann plot for the $2\nu_1$ band in the data set).
The two P(11) transitions  for both $2\nu_1$ and $\nu_2+\nu_3+2\nu_5^0$ bands depicted in Figure~\ref{fig1} were 
obtained with 22 thousand and 39 thousand of primary \HCNHp\ ions, respectively. 
Additionally, taking into account the actual laser power, the resulting estimate for intensity ratio from the product ion signal
between these two transitions is 25:1. 
This is in very good agreement with the calculated ratio of 31:1 (see Table \ref{t_constants} and 
section \ref{Computational} for details) and showcases the very high sensitivity of the ``kick-out'' LIR technique.
The strongest fundamental band $\nu_1$ (NH stretch) of \ce{HCNH+} has
a calculated intensity of $482.46\;\kmmol$,
almost four orders of magnitude stronger,
than that of the weak combination band, $0.06\;\kmmol$. 
This intensity translates into a transition dipole moment of only $1.9\cdot10^{-3}$ Debye and to our knowledge it
is the first detection of a transition involving simultaneous excitation of 4 quanta in a high resolution ion trap based 
action spectroscopy experiment.
We are aware of only one comparable ion trap experiment involving 3 quanta, the $3\nu$ R(0) \ce{OH-} transition studied in \cite{Lakhmanskaya2020}.

Previous theoretical studies \cite{Botschwina1994,Peterson1995,Cotton2013} focused on fundamental transitions or covered 
only the lower lying vibrational states \cite{Brites2012}. 
\citet{Botschwina1986} arbitrarily corrected calculated potential energy curves to exactly reproduce the experimental band origins 
for $\nu_1$ and $\nu_2$. 
The resulting predicted band origins for $2\nu_1$ (6858~\rcm), $\nu_1+\nu_2$ (6640~\rcm) and $2\nu_2$ (6281~\rcm) are very 
close to the present experimental values of 6846.77981(90)~\rcm, 6640.47624(43)~\rcm\ and 6282.03578(63)~\rcm, respectively.

\section{Conclusion}\label{}
Band origins and upper state spectroscopic constants for four new overtone and
combination bands of \HCNHp\ ions were determined using a ``kick-out'' enhanced
multi-step laser induced reaction technique in a 22 pole radio frequency ion trap. 
The application of zero background spectroscopy enables the observation of
very weak transitions such as the $\nu_2+\nu_3+2\nu_5^0$ reported in this study.
The provided state-of-the-art \emph{ab initio} calculations
are in good agreement with the determined band origins,
despite the fact that computing highly excited vibrational states
(overtones and combination bands) with high accuracy can be much more challenging.
Using these frequencies, the \ce{HCNH+} ion can now easily be experimentally monitored
in optically thin environments using cheap semiconductor distributed-feedback (DFB)
diode lasers in the S, C, L telecommunication bands or in emission using spectrometers.

\section*{Declaration of Competing Interest}
The authors report there are no competing interests to declare.

\section*{Data Availability}
The data that support the findings of this study
(raw experimental data, as well as the calculated PES) 
are openly available in Zenodo at
\href{https://doi.org/10.5281/zenodo.12794341}{10.5281/zenodo.12794341},
reference number 12794341.

\section*{Acknowledgments}

This work was supported by the Max Planck Society; 
Czech Science Foundation projects GACR 22-05935S, 23-05439S, 24-10992S. 
M.K. acknowledges support by Charles University project GAUK 337821 and that 
this article is based upon work from COST Action CA21101, supported by COST (European Cooperation in Science and Technology).
J.S. has been supported by an ERC Advanced Grant (MissIons: 101020583).
We thank Prof. Stephan Schlemmer (Uni. zu K\"{o}ln) for lending of the Agilent laser system.
We thank the reviewers for their constructive feedback.

\appendix
\setcounter{table}{0}
\renewcommand*\thetable{\Alph{section}.\arabic{table}}

\section{Line lists of measured \ce{HCNH+} transitions}

\begin{table*}[width=0.9\textwidth,cols=6,pos=h]
    \caption[]{Experimentally measured transitions of \HCNHp.}\label{t_lines}
    \begin{tabular*}{\tblwidth}{@{} llllllll@{} }
        \toprule
        \multicolumn{2}{c}{$2\nu_1$} & \multicolumn{2}{c}{$\nu_1 + \nu_2$} & \multicolumn{2}{c}{$2\nu_2$} & 
            \multicolumn{2}{c}{$\nu_2 + \nu_3 + 2\nu_5^0$} \\
        \midrule
        R(14) & 6880.63603(26)  & R(9)  & 6663.59689(6)   & R(7) & 6300.70083(12)   & R(7)  & 6607.47736(39)   \\
        R(13) & 6878.57219(20)  & R(7)  & 6659.20677(7)   & R(6) & 6298.47618(11)   & R(6)  & 6605.18607(25)   \\
        R(12) & 6876.48002(17)  & R(5)  & 6654.69970(6)   & R(5) & 6296.21932(24)   & P(4)  & 6578.47191(37)   \\
        R(11) & 6874.36157(14)  & P(4)  & 6630.41455(7)   & R(4) & 6293.93282(24)   & P(6)  & 6573.33680(39)   \\
        R(10) & 6872.21087(12)  & P(5)  & 6627.82652(5)   & R(3) & 6291.61514(11)   & P(8)  & 6568.10975(33)   \\
        R(8)  & 6867.83104(10)  & P(7)  & 6622.56541(9)   & R(1) & 6286.88797(21)   & P(10) & 6562.78773(27)   \\
        R(7)  & 6865.60037(16)  & P(9)  & 6617.19029(20)  & R(0) & 6284.47692(32)   & P(11) & 6560.09040(32)   \\
        P(3)  & 6839.28406(11)  & P(11) & 6611.69643(12)  &      &                  & P(12) & 6557.37146(48)   \\
        P(4)  & 6836.73553(10)  & P(12) & 6608.90821(14)  &      &                  &       &                  \\
        P(5)  & 6834.15802(7)   & P(13) & 6606.09246(20)  &      &                  &       &                  \\
        P(6)  & 6831.55345(6)   & P(14) & 6603.24654(10)  &      &                  &       &                  \\
        P(7)  & 6828.92488(10)  &       &                 &      &                  &       &                  \\
        P(8)  & 6826.26870(7)   &       &                 &      &                  &       &                  \\
        P(9)  & 6823.58547(5)   &       &                 &      &                  &       &                  \\
        P(10) & 6820.87529(5)   &       &                 &      &                  &       &                  \\
        P(11) & 6818.13767(6)   &       &                 &      &                  &       &                  \\
        P(13) & 6812.58447(21)  &       &                 &      &                  &       &                  \\
        \noalign{\smallskip}
        \bottomrule
    \end{tabular*}
    \vskip 0.2em
    {\footnotesize{\emph{Note:}
    All values in units of \rcm. In all cases the lower state is the vibrational ground state. 
    Numbers in parentheses are statistical errors of the fit in units of the last quoted digit.
    }}
\end{table*}

\bibliographystyle{elsarticle-num-names}

%\bibliography{hcnh}

\end{document}